\documentclass{ws-ijmpd}
\usepackage{amsmath,amssymb,amsfonts}
\usepackage{color}
\usepackage{amsfonts}
\usepackage{amssymb}
\usepackage{slashed}
\usepackage{graphicx}
\usepackage{tikz}
\usetikzlibrary{positioning}
\usetikzlibrary{decorations.pathmorphing}
\usetikzlibrary{decorations.markings}
\usetikzlibrary{snakes}
\usepackage{float}
\usepackage{graphicx}%
\newcommand{\be}{\begin{equation}}
\newcommand{\ba}{\begin{eqnarray}}
\newcommand{\bal}{\begin{align}}
\newcommand{\eal}{\end{align}}
\newcommand{\ee}{\end{equation}}
\newcommand{\ea}{\end{eqnarray}}

\def\VEV#1{{\left\langle #1 \right\rangle}}

\def\Pl{{\rm Pl}}
\def\hp{h_+}
\def\hc{h_{\times}}

\def\yboxit#1#2{\vbox{\hrule height #1 \hbox{\vrule width #1
    \vbox{#2}\vrule width #1 }\hrule height #1 }}
\def\fillbox#1{\hbox to #1{\vbox to #1{\vfil}\hfil}}
\def\ybox{\yboxit{0.4pt}{\fillbox{8pt}}\hskip-0.4pt}
\def\VEV#1{\langle{ #1} \rangle}

\def\hp{h_+}
\def\hc{h_{\times}}
\def\yboxit#1#2{\vbox{\hrule height #1 \hbox{\vrule width #1
    \vbox{#2}\vrule width #1 }\hrule height #1 }}
\def\fillbox#1{\hbox to #1{\vbox to #1{\vfil}\hfil}}
\def\ybox{\yboxit{0.4pt}{\fillbox{8pt}}\hskip-0.4pt}

\newcommand{\bea}{\begin{eqnarray}}
\newcommand{\eea}{\end{eqnarray}}
\newcommand{\bse}{\begin{subequations}}
\newcommand{\ese}{\end{subequations}}
\newcommand{\barr}{\begin{array}}
\newcommand{\earr}{\end{array}}

\newcommand{\lsim}{\lesssim}
\newcommand{\gsim}{\gtrsim}

\def\tt{t\bar t}

\def\pslash{\not{\hbox{\kern-4pt $p$}}}
\def\qslash{\not{\hbox{\kern-4pt $q$}}}
\def\lv{\not{\hbox{\kern-4pt $L$}}}
\def\lsim{\mathrel{\raise.3ex\hbox{$<$\kern-.75em\lower1ex\hbox{$\sim$}}}}
\def\gsim{\mathrel{\raise.3ex\hbox{$>$\kern-.75em\lower1ex\hbox{$\sim$}}}}
\def\ifmath#1{\relax\ifmmode #1\else $#1$\fi}

\let\<=\langle
\let\>=\rangle

\let\+=\uparrow

\usepackage[super,compress]{cite}
\begin{document}


%
\catchline{}{}{}{}{}
%

\title{INFLATIONARY BIREFRINGENCE AND BARYOGENESIS }

\author{STEPHON ALEXANDER}

\address{Brown University, Department of Physics\\
Providence, RI 02912,
USA\\
stephon\_alexander@brown.edu}

\maketitle


\begin{abstract}
A decade ago, the first leptogenesis model based on inflation was proposed, where the complex phase of the inflaton field carries lepton number\cite{Alexander:2004us}.  If the inflaton field is an axion, it can couple to gravitational waves and gauge fields via. Chern-Simons invariants.  Due to these couplings, birefringent gravitational and gauge primordial perturbations are created during inflation to generate a lepton asymmetry, establishing a possible connection between non-vanishing TB-parity violating polarization cross-correlations and leptogenesis.  We also discuss the prospect for a subset of these models can directly source circular (V-mode) polarization in the CMB.
\end{abstract}



\section{Introduction}
It remains a challenge to find a testable mechanism of the observed baryon asymmetry in the universe.   Not only is it important to explain why there is an asymmetry, but also the tiny number associated with it, the baryon asymmetry index $\frac{n_B}{n_\gamma} = (6.5\pm 0.4)\times 10^{-10}\ $.   This number is also crucial for establishing the abundance of light elements in galaxies.  Sakharov spelled out the necessary conditions to obtain the baryon asymmetry and many physicists sought models that occured either at the end of inflation or much later.

The quantum fluctuations generated during inflation acts as the seeds for large scale structure, but had very little to say about the matter-antimatter asymmetry.  In 2005, Peskin, Sheikh-Jabbari and the author wondered if leptogenesis could happen during inflation and were successful in constructing the first in a series of such models\cite{Alexander:2004us}. Prior to that time, it was thought that the exponential dilution of matter would win over any particle production during inflation.  But there was an interesting loophole that forged the way to progress: If the inflaton carries lepton number then the inflaton's amplifcation during cosmic inflation may counter the dilution of leptons in an expanding background.

The key to all instantiations of inflationary baryogenesis is for the inflaton field to simultaneously carry the lepton or baryon number and induce CP violation\cite{Alexander:2004us}.  During inflation the VEV of the inflaton field will not redshift since it is slowly rolling and its superhorizon quantum fluctuations are also amplified.  Subsequently, this VEV will continually source lepton/baryon number despite the rapid expansion rate.  The inflaton field can carry the lepton charge in two ways.  First, via an Affleck-Dine like mechanism there are scalar SUSY superpartners of quarks and leptons called squarks and sleptons that carry B-L charge via $n_{B-L} = 2qIm[\phi^{*}\dot{\phi}]$\cite{Hertzberg}.  This idea was used by Hertzberg and Karouby to generate a baryon asymmetry during inflation.  Another way that the inflaton field can carry the lepton number is as a pseudoscalar with an anomalous or direct coupling to the baryon/lepton current\cite{Alexander:2004us,adshead}.  In this case the pseudoscalar is associated with the phase of the lepton/baryon charge, and couples to gauge and gravitational global anomalies.   As the inflaton rolls it acts as source for birefreingent gauge and gravitional waves, which provides another way to source the lepton/baryon asymmetry.  In what follows I will spell out the general pillars for inflationary birefringence and baryogenesis (IBB) and specify on the case of parity violating gauge and gravitational waves.

\label{sec:intro}
\noindent

\section{Inflationary Leptogenesis and Birefringence}
\noindent

Leptogenesis is a valid route to explain the baryon asymmetry, provided 
that the Sakharov conditions for leptons are satisfied:
\begin{enumerate}
\item lepton number violating interactions should be operational.
\item Charge and parity (CP) should be violated. 
\item CP and baryon number violating interactions should be active when the Universe is out of thermal equilibrium.
\end{enumerate}

Once leptogenesis is established, the weak interaction contain processes,
mediated by {\it sphalerons} ($SU(2)$ instantons), which converts leptons into baryons and are
thermally activated at temperatures greater than 1TeV.  Therefore, a baryon asymmetry can arise 
through the generation of net lepton number at high temperature
through out-of-equilibrium and CP-asymmetric processes. 

A leptogenesis mechanism that satisfies all three Sakharov conditions was first realized if the inflaton field is associated with a complex
modulus field.  The simplest model of this kind is that of pseudo-scalar $\phi$ as the inflaton\cite{Alexander:2004us}.  In this case the pseudo-scalar VEV carries the lepton number through an anamalous coupling to the global lepton gravitational anomaly \cite{Alex,Pi,Lue,AY,shahin}. 

During inflation the slow rolling of the inflaton will source parity violating gravitational waves.  The specific form of the gravitational waves will generate a non-vanishing lepton current via. a gravitational anomaly in the standard model.  Explicitly, due to the gravitational triangle anomaly, the lepton current is
\be\label{Jlepton}
     \partial_\mu J^\mu_\ell  =  \frac{N}{32\pi^2}   \epsilon^{\alpha\beta\gamma\delta}
 R_{\alpha\beta \rho\sigma} R_{\gamma\delta}{}^{\rho\sigma}
\ee
where
\be
J^{\mu}_{\ell} =   \sum_{i=L,R} \bar \ell_i\gamma^\mu \ell_i + \bar \nu_i \gamma^\mu \nu_i 
\ee

We aim to show that the inflaton field can source a non-vanishing lepton number via. birefringent gravitational waves that have a non-vanishing anomaly.  We begin with the following action:


\begin{eqnarray} \label{action1}
\!\!\!\!S= \!\!\int_{\mathcal{M}_4}\!\!\!\!\!\! d^4x\sqrt{-g} \Bigg[\frac{M_p^2\,R}{8\pi} \!-\! \frac{1}{2}\partial_{\mu}\phi\,\partial^{\mu}\phi \!-\! V(\phi) +\! F(\phi)\epsilon^{\alpha\beta\gamma\delta}
 R_{\alpha\beta \rho\sigma} R_{\gamma\delta}{}^{\rho\sigma} \Bigg]\,,
\end{eqnarray}

After linearizing the gravitational sector of the action, working in TT gauge, we get the following Lagrangian:

\be\label{RRdual2}
\begin{split}
{\cal L}&= -({h_L} \Box {h_R} + {h_R} \Box {h_L})\cr
& +{16i F(\phi)}\biggl[\left(\frac{\partial^2}{\partial z^2}{h_R}
\frac{\partial^2}{\partial t\partial z}{h_L} -
\frac{\partial^2}{\partial z^2}{h_L}
\frac{\partial^2}{\partial t\partial z}{h_R} \right)\cr
& + a^2 \left(\frac{\partial^2}{\partial t^2}{h_R}
\frac{\partial^2}{\partial t\partial z}{h_L} -
\frac{\partial^2}{\partial t^2}{h_L}
\frac{\partial^2}{\partial t\partial z}{h_R} \right) \cr
&  + Ha^2\left(
\frac{\partial}{\partial t}{h_R}\frac{\partial^2}{\partial t\partial z}{h_L}
-\frac{\partial}{\partial t}{h_L}\frac{\partial^2}{\partial t\partial
z}{h_R}\right)\biggr] +{\cal O}(h^4)
\end{split}
\ee

\subsection{Birefringent Gravitational Waves and Parity Violation}

To see the parity violation more explicitly, it is convenient to define a
helicity basis in terms of the transverse and traceless elements of the tensor perturbation $h_{ij}$.
\be
h_L =  (\hp - i \hc){/\sqrt{2}} \ , \qquad
h_R = (\hp + i \hc){/\sqrt{2}} \ .
\ee
Here $h_L$ and $h_R$ are complex conjugate scalar fields. To be very explicit,
the negative frequency part of $h_L$ is the conjugate of the positive
frequency part of $h_R$.  It is straightforward to obtain equations of motion for the left and right handed gravitational waves $h_L$ and $h_R$ from (\ref{RRdual2}):
\be\label{LReqs}
  \ybox\, h_L = - 2i \frac{\Theta}{ a} {\dot h}^\prime_L \ , \qquad
  \ybox\, h_R = + 2i \frac{\Theta}{a} {\dot h}^\prime_R \ ,
\ee
where
\be\label{Thetaval}
\begin{split}
  \Theta &\simeq 4(F''\dot\phi^2+2F'H\dot\phi)/M^2_{Pl} \\
\end{split}
\ee

In what follows, we will consider a linear dependence on the pseudoscalar $F(\phi) = \frac{N}{16\pi^{2}M_{Pl}^{2}}\frac{\phi}{M_{Pl}}$, where $N$ is the number of fermions in the triangle loop in the evaluation of the Chiral-Anomaly.  This linear dependence was demonstrated to consistent with the consistency relations in slow roll inflation \cite{Alexander:2004wk}.

The solutions are found to be of a plane-wave form whose amplitude grows in time and proportional to the velocity of the inflaton

\be\label{gdef}
     h_L = e^{ikz} \cdot (-ik\eta)
                        e^{k\Theta\eta}  g(\eta)
\ee

where 
\be\label{findg}
    g(\eta) =   \exp[ ik(1-\Theta^2)^{1/2} \eta (1 + \alpha(\eta))] \ ,
\ee
where $\alpha(\eta) \sim \log \eta/\eta$.

The right handed gravity wave will have the same solution whose exponent will have the opposite sign to $h_{L}$, reflecting parity violation.  Given these solutions we can directly compute the graviational anomaly:

\be\label{RRdual}
\begin{split}
R\tilde R= \frac{4i}{a^{3}}\biggl[&\bigg(
{\partial^2_z} h_R\ {\partial_{z}\partial_{t}}h_L+a^2 {\partial^2_{t}}h_R\
{\partial_{t}\partial_{z}}h_L\cr
& +\frac{1}{2}{\partial_t}a^2 {\partial_t} h_R\
{\partial_{t}\partial_{z}}h_L\bigg)
-{\left(L\leftrightarrow R\right)}
\biggr]
\end{split}
\ee
We find 
\be\label{RRdualval}
  \VEV{R\tilde R} =  {16\over a^4}\, \int \, {d^3 k\over (2\pi)^3}\
  {H^2\over 2 k^3 M_\Pl^2}
                      \cdot k^4 \Theta  + {\cal O}(\Theta^3)
\ee
where we have picked up only the leading behavior for  $k\eta \gg 1$.
The integral over $\eta$ is dominated at large values of
$\eta$, early times. We are now ready to evaluate the lepton density that arises through the gravitational
anomaly \eqref{Jlepton}. Inserting \eqref{RRdualval} into \eqref{Jlepton} and
integrating over the time period of inflation, we obtain
\be\label{netlept}
  n = \int^{H^{-1}}_0 d\eta\ \int \, {d^3 k\over (2\pi)^3}\
       {3\over 16\pi^2}\, {16 H^2  k  \Theta \over M_\Pl^2}  \,,
\ee
where $n$ is the lepton number density.  The time integral represents a compromise between
two effects of inflation, first, to blow up distances
and thus carry us to smaller physical momenta and, second,
to dilute the generated lepton number through expansion.
The integral over $k$ runs over all of momentum space, up to the
scale $\mu$ at which our effective Lagrangian description breaks down.  
After performing the $k$ integral we obtain,
\be\label{nval}
 n  =  {N\over 48\pi^4} \left({ H\over
            M_\Pl}\right)^2 \Theta H^3  \left({\mu\over H} \right)^4 \ .
\ee 

Remarkably this expression contains some interesting physics. First, the factor
$(H/M_\Pl)^2$ is the  magnitude of the gravity wave power
spectrum. We should emphasize that the typical gravity wave power
spectrum comes from the super-horizon modes, while the main
contribution to $n$ comes from the sub-horizon modes. Second, the
factor $\Theta$ is  a measure of the effective CP violation caused by birefringent
gravity waves.  Finally, the factor
$({\mu/H})^4$ gives the UV enhancement due to the short distance fluctuations to
generate $R\tilde R$, rather than the super-horizon modes.

We can use the Friedmann equations to get the entropy density $s$, which is $s =  2\pi^2 g_* T^3/45$, where $T$ is the reheating temperature, $g_*$ is the effective
number of massless degrees of freedom.  We finally arrive at the baryon asymmetry index
\be\label{n/n-final}
\frac{n_B}{n_\gamma} \simeq 1.3\times 10^{-5} g_*^{3/4}\sqrt{\epsilon}\ {\cal N} \left(\frac{H}{M_{Pl}}\right)^{3/2}
\left(\frac{\mu}{M_{Pl}}\right)^4  .
\ee
These birefringent gravitational waves are also generated on superhorizon scales and can lead to non-vanishing TB cross-correlation in the CMB polarization powerspectrum \cite{Alexander:2004wk,contaldi,Alex}
Within the usual supersymmetric particle physics models $g_* \sim 1000$ is a reasonable choice.
The WMAP data, through the density perturbation ratio $\delta\rho/\rho$ (for a single field inflation)
leads to an upper bound on $H/M_{Pl}$ ratio to be $H\lesssim 10^{14}\ GeV$.

The factor ${\cal N}$ is inferred from the string theory compactification and is proportional to
the square of the four dimensional $M_{Pl}$ to ten dimensional (fundamental) Planck mass \cite{A-G}.   It was later shown that gravitational Chern-Simons coupling can also give a sufficient amount of preheating through parametric resonance of the birefringent gravitational waves at the end of inflation \cite{cormack}. 

\section{Baryogenesis from Hypercharge Anomaly}

Gravitational waves are not the only way to generate baryogenesis during inflation.  In follow up work the authors \cite{AMS} showed that the baryogenesis can occur from a coupling between the axion and a gauge Chern-Simons term associated with the $U(1)_{Y}$ hypercharge sector of the standard model, $\delta L = \frac{\alpha}{f} \phi Y\tilde Y$.  From this term birefringent Hypercharge perturbations will be produced since their equations of motion will be modified\footnote{Various authors have discussed parity violation of gauge fields coupled to the Chern-Simons term as well as non-gaussianity \cite{Sorbo1,Sorbo2,Peloso}} :
\be \label{ser}
\delta\ddot{Y} (\eta, \vec{k})_h + \left[ k^2 + 2 \, h \, \frac{\xi}{\eta}   \right] \delta Y(\eta, \vec{k})_h =0\,,
\ee
where  $\xi = \frac{\alpha\dot{\phi}}{f}$ and $h=\pm1$ is the helicity.  So similar to gravitational waves, opposite helicity gauge fields will have different dispersion relations.   This birefringent hypercarge will generate a hypermagnetic field which will source a net right handed electrons due to the Hypercharge anomaly \cite{Bamba:2007hf}:
 
\be \label{chiral}
\nabla_{\mu} \mathcal{J}^{\mu}_{e_{R}}= 
\frac{y_{R}}{32 \pi^2}Y_{\alpha \beta} Y_{\mu \nu} \epsilon^{\alpha \beta \mu \nu}\,,
\ee
where $Y_{\mu\nu}$ is the hypercharge field strength, $y_{R}=-2$ is the right handed electron hypercharge and $\cal{J}\rm^{\mu}_{e_{R}}$ is the right-handed electron current.  Anber et. al. \cite{Anber} demonstrated that a baryon overproduction occurs with most of the parameter space and of the pseudoscalar inflation as well as accounting for the conductivity of the plasma during reheating.  The backreaction of these birefringent gauge fields also leads to parity violating gravitational waves, providing another window to connect both models of baryogenesis \cite{Peloso,Cook,Alexander}

\section{Conclusion and Discussion}

A detection of cosmic birefringence in the polarization powerspectrum is an excting undertaking both from observational and experimental fronts\cite{Yadav,Ade}.  In this talk, I discussed the possibility that a detection of a TB powerspectrum can be linked with baryogenesis via. birefringent gauge or gravitational perturbations that are produced during inflation.  It is intriguing that the inflaton field not only provides the properties to address the problems of the standard big bang, but may also be the agent responsible for the cosmic baryon asymmetry.   It is worthwhile to ask if there are other ways to experimentally probe and distinguish different models of inflationary baryogenesis.  Recent work by Contaldi, Magueijo and Smolin considered similar parity violating gravitational waves where the amplitudes between different handedness is captured by an asymmetry parameter $\gamma$ \cite{Contaldi:2008yz}.  These authors found that the $BB$ quadrapole power of the polarization cross correlation will be smaller than the parity violating $TB$ cross-correlation by \be \frac{C^{TB}_{2}}{C^{TT}_{2}} \sim 1\times 10^{-3}\frac{r}{\gamma}. \ee  Here $\gamma$ captures the difference in the amplitude between left and right handed gravitational waves and $r$ is the tensor to scalar ratio.  In our model we saw that the sub-horizon gravitational wave modes contributed to the lepton asymmetry.  While we expect a similar modification in the $TB$ amplitude of the superhorizon modes considered by Contaldi et al., we leave it up to future work to see precisely how much $TB$ cross-correlation amplitude birefringence is attained in specific models.  The author and his collaborators are currently investigating this possibiity as well as the potential of certain birefringence models to produce a direct source of circular (V) polarization. \cite{Alex2,BGA}  
\section*{Acknowledgments}
I thank my collaborators, Michael Peskin, S. James Gates, Antonino Marciano, David Spergel, Nico Yunes, Brian Keating, Shahin Sheikh-Jabbari and Jerome Martin for the years of collaboration on the ideas presented in this talk. This work is supported in part by DOE Grant No. DESC0010386.


\end{document}